\documentclass{aa}
\pdfoutput=1

\usepackage{ifpdf,epsfig,latexsym,longtable,lscape,supertabular}
\usepackage[authoryear]{natbib}

\def\e20{$\times 10^{20}$}
\def\ergsec{erg s$^{-1}$}
\def\ergcmsec{erg cm$^{-2}$ s$^{-1}$}

\def\today{\number\day -\number\month -\number\year}

\def\ii{\,{\sc i}\,{\sc i}\,}
\def\hi{H\,{\sc i}}
\def\scg{SCG0018-4854}

\begin{document}
\pagenumbering{arabic}
\title{Detection of a hot intergalactic medium in the spiral-only
compact group SCG0018-4854 }

\author{
G. Trinchieri\inst{1}
\and
A. Iovino\inst{1}
\and
E. Pompei\inst{2}
\and
M. Dahlem\inst{3}
\and
J. Reeves\inst{4}
\and
R. Coziol\inst{5}
\and
S. Temporin\inst{6}}
\institute{
INAF-Osservatorio Astronomico di Brera, via Brera 28, 20121
 Milano Italy\\
 \email{ginevra.trinchieri@brera.inaf.it}
 \and
ESO - Alonso de Cordova 3107 Vitacura Casilla 19001 Santiago 19, Chile
\and
CSIRO/ATNF - Paul Wild Observatory Narrabri NSW 2390 AUSTRALIA 
\and
Astrophysics Group, School of Physical and Geographical Sciences, Keele
University, Keele, Staffordshire ST5 5BG, UK.
\and
Departamento de Astronomía, Universidad de Guanajuato, 36000 Guanajuato, Guanajuato, Mexico
\and
Laboratoire AIM, CEA/DSM - CNRS - Université Paris Diderot,
DAPNIA/SAp, 91191 Gif sur Yvette, France
}

   \date{Draft: \today}


\abstract
  {}
  {Compact groups of galaxies are excellent laboratories
for studying galaxy interactions and their effects on the evolution
of galaxies. In particular, dynamically young systems, with a large
fraction of interacting, late type galaxies, have so far escaped proper
studies in the X-ray band and their hot intergalactic medium properties
are virtually unknown. Motivated by this lack of knowledge, we present
a detailed investigation here 
of the X-ray properties of such a dynamically young system.}
  {We obtained XMM-Newton observations of one spiral-only system in
  the new southern compact group catalogue: \scg.  We present here the
  results of the data analysis and discuss them  in comparison  with the 
  few other similar systems also studied in the X-ray band.}
  {The 4 members of \scg\ emit at a level
  comparable to what is expected based on their optical properties. We 
  detect the low level diffuse emission between galaxies, which is most
  likely due to the presence of the intergalactic medium.  Compared to other
  spiral--only groups, \scg\ could be the coolest
  system detected so far, although the measurements have
  large uncertainties. These results indicate that high quality deep 
  X-ray observations are needed to allow a proper study of 
  the properties of the potential well of dynamically young systems, 
  before more general conclusions can be drawn on their average characteristics.}
  {}

\keywords{ISM: general; X-rays: galaxies: clusters; Galaxies: ISM;
X-rays: ISM}

\titlerunning{ Hot IGM in \scg}

\maketitle

\section{Introduction}

Compact groups (CGs) are small associations of galaxies (4-8)
characterised by high over-densities ($\delta\rho$/$\rho$ $\sim$ 80)
and low velocity dispersions, in the range 13 to 617 km s$^{-1}$, 
with a median value of 200 km s$^{-1}$ \citep{Hickson92}.  It
seems now well established that these structures are physically real,
and that they are likely to play an important role in the environmentally driven
evolution of galaxies \citep[see e.g.][]{MendesHickson94, CoziolPlauchu-Frayn}.

Recent optical results \citep{coziol} indicate that we observe CGs 
with a wide range of properties. In some of them,  member galaxies 
show significant amounts of ionised gas in their nuclei, indicating 
recent star formation or low luminosity AGNs.  Others are instead 
characterised by a low level of activity of their members, which are
mainly E/S0 galaxies with old stellar populations. In a fraction of 
CGs an intermediate age stellar population is present in the member 
galaxies, consistent with a post-starburst phase.

A reasonable evolutionary sequence would see CGs evolve from
dynamically young groups, expected to be dominated by a population
of late-type galaxies, to more relaxed systems, 
dominated by an early-type galaxy
population, through a multiplicity of mechanisms such as mergers of
member galaxies, accretion of new gas from external reservoirs, infall
of new galaxies and heating of the intra-group medium through
dynamical friction and AGN feedback \citep{coziol04,gomez}.

Inserting this evolution in the larger picture of hierarchical
structure formation theory, one would expect that CGs, being the lower
mass tail of structures in the Universe, would have formed more
recently than eg. massive cluster of galaxies \citep{coziol04}. This would be
particularly true for dynamically young systems, that in the evolutionary sequence above
should be at the lowest mass tail of CGs.

The observations of X-ray emission in CGs should be interpreted within
this context. In the early discussions about the nature of CGs, the
presence of a hot intergalactic gas in these systems was taken as
evidence that these are physically bound structures with a common
potential well. 

The presence of hot inter galactic gas in CGs was confirmed by \citet{Ponman}, and it 
was further established that its characteristics 
are indistinguishable from those of the more general class of groups \citep{HelsdonPonman}. 
Taken as a whole, groups form the fainter part of the
X-ray luminosity function or luminosity vs.\ velocity dispersion
relation for clusters \citep{Ponman, HelsdonPonman, Mul03, Jeltema}, a result fully
consistent with the predictions of the hierarchical structure
formation theory.

Another interesting matter concerning X-ray emission in CGs is the
problem of the so called {\it fossil groups} \citep{Ponman94,
Vikhlinin99,jones03}. These groups have been suggested as the possible
end product of the evolution of CGs. However some observational
differences between compact groups and fossil groups seem to argue
against this hypothesis.  Among them, a possible difference in mass
range between fossil groups and compact groups, the former so far
observed at the higher end of the mass distribution
\cite[10$^{13}$-10$^{14}$ vs 10$^{11}$-10$^{14}$ M$_{\sun}$, see,
e.g.][]{jones03, Hickson92, Pompei07}, and in X-ray luminosity, where
again fossil groups are observed at the bright end \citep[L$_x >$
10$^{42}$ vs 10$^{41}$-10$^{42}$ erg s$^{-1}$][]{jones03}.  This,
however could just be more an observational bias than a real problem
as systematic searches for low X-ray/mass fossil groups are still
lacking. Furthermore in the hierarchical structure formation model
fossil groups could be the end result of the massive groups formed at
early cosmic times that coalesced in the single massive X-ray bright
galaxy observed today \citep[see eg.][]{Benda}. 

So far very little is known about the X-ray properties of
spiral-only compact groups, which would presumably be at the
beginning of the simple evolutionary sequence presented above. This
is due to the small fraction of such systems in the best-studied
catalogue of CGs by \citet{Hickson} and to the difficulty in
observing them to the appropriate sensitivity levels in the X-ray
band. To date, a hot IGM has been detected only in a few spiral
dominated groups. The only example in Hickson's catalog is HCG 16
\citep{Bel03}, which is also a quite remarkably active group \citep{ribeiro}. A few
other cases of diffuse X-ray emission in spiral-dominated groups
have also been recently reported in a higher redshift sample
\citep{Mulchaey06}.

X-ray emission in these systems can be an important probe both of the
dynamical (by probing the system's potential), and activity stage (by
completing the information on galaxy activity) of the group itself.
Can we distinguish between young and evolved CGs based on their X-ray
emission? In particular, what is the origin of the intergalactic gas
in CGs? According to the hierarchical structures formation model we
would expect recently formed CGs to be relatively poor in hot gas, due
to their smaller potential wells.

In this paper we present the hot diffuse component and the high
energy properties of one such group, \scg, selected from the sample
of Southern Compact Groups \citep[SCGs,][]{Iovino2000,Iovino03}. The
SCGs is a highly complete sample of CGs, selected in an automated
fashion from a galaxy catalog obtained from UKST digitised blue
plates. Follow-up spectroscopic observations of the brightest
candidates (with a blue magnitude of the brightest galaxy of
$<14.5$) have confirmed the CG structure, with three or more
concordant member redshifts, for 49 of the 60 pre-selected
candidates. Preliminary studies of the atomic gas content of these
systems have just begun with a pilot sample of six \citep{Pompei07}.

\scg\ is a beautiful tight quartet of galaxies, already listed in
the Rose compilation of CGs as Rose34 \citep{Rose77}.  It is located
at a recession velocity of $\sim 3300$ km/sec and consists of four
galaxies (NGC~88, NGC~89, NGC~87 and NGC~92) in an extremely tight
configuration on the sky: all within a circle of radius of
1$\farcm$6. A possible fifth member (ESO 194-G 013) is  at a larger
projected distance of $\sim 15'$ (at the distance of the group $1'
\sim 15\ h_{70}^{-1}$ kpc). The velocity dispersion of only 120 km
s$^{-1}$ is very low, albeit with a large associated uncertainty
because it is derived from only five individual redshift
measurements. All members of the tight quartet exhibit a disturbed
morphology, including a spectacular extended tidal tail of the
brightest galaxy \cite[NCG~92, see also][]{Temporin, Pompei07}.

All these observational properties support that \scg\ is one good
example of a ``dynamically young'' CG and, as such, a particularly
interesting target for the XMM-Newton observations presented here.

We adopt $H_{0}$ = 70 km s$^{-1}$ Mpc$^{-1}$ throughout.

\section{Observation and Data Reduction}

\begin{table*}
\caption{Journal of the XMM-Newton EPIC observations.}
\begin{tabular}{lccccccll}
\hline
 \multicolumn{1}{l}{Observation} & 
 \multicolumn{3}{c}{Original Exp. Time (ks)} &
  \multicolumn{3}{c}{Cleaned Exp. Time (ks)} &
  \multicolumn{1}{l}{RA$^\dagger$} &
  \multicolumn{1}{l}{DEC$^\dagger$} \\
    \multicolumn{1}{c}{ Id} &
 \multicolumn{3}{c}{EPIC-pn/MOS1/MOS2}
 & \multicolumn{3}{c}{EPIC-pn/MOS1/MOS2}
 & \multicolumn{2}{c}{J2000} \\
\hline
0152330101 & 35.3& 39.2 & 39.3 & 16.7& 29.9& 29.9&  00$^h$21$^m 23\fs30$
& -48\degr38$'23\farcs8$\\
\hline
\end{tabular}
\label{log}

\noindent $^\dagger$ Position of the target
\end{table*}

\begin{figure*}
\resizebox{18cm}{!}{
\psfig{figure=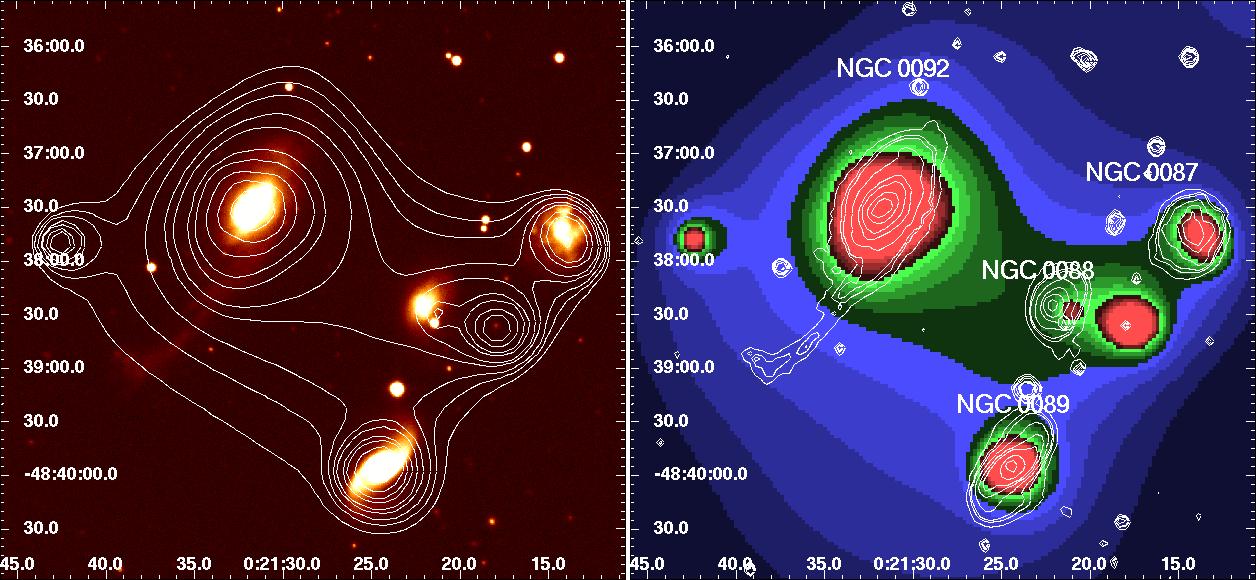,width=18cm}
}
\caption{LEFT: Iso-intensity contours of the emission in the 0.3-2.0
keV energy band from the EPIC-MOS data overlayed onto a 300s V band
image obtained with DFOSC at the 1.54mt Danish Telescope.  An
adaptively smooth algorithm is applied.  
Contours are at: 0.033, 0.035, 0.038, 0.043, 0.048, 0.063, 0.078, 0.125,
0.238, 0.75, 1.625 cnt $arcsec^{-2}$.  
RIGHT: The optical contours
are overlayed onto the X-ray smoothed image, displayed in logarithmic
scale.}
\label{figure1} 
\end{figure*}

We obtained a $\sim 35$ ks XMM-Newton \citep{refxmm}
observation of SGC0018-4845 
with EPIC with its medium filter in December 2002 (see Table~\ref{log}).
We used the XMM-Newton Science Analysis System (SAS, version
$xmmsas\_20060628\_1801$) to update the event files with the newest
calibration and to clean it from flaring events (see the science
threads at {\tt http://xmm.vilspa.esa.es}).  A significant increase of
the background is observed towards the end of the observation, that
had to be excluded from the analysis.  Additional flaring episodes are
observed throughout the observation.  Cleaning the data from all high
background episodes reduces the net observing time to $\sim 16.7 -
29.9 $ ks for EPIC-pn and EPIC-MOS, respectively.  We retain single, double 
and quadruple events for the morphological analysis, and for the MOS
spectral data.  For the spectral analysis of the EPIC-pn data we only
consider single and double events.  We also use both CIAO ({\tt
http://cxc.harvard.edu/ciao/}) and DS9-Funtools ({\tt
http://head-cfa.harvard.edu/RD}) astronomical software to analyse the
data and present results.

To improve the photon statistics we have summed all EPIC-MOS data for 
the spatial analysis.  The  EPIC-pn data are kept separate to avoid 
artifacts that could be introduced by the different patterns in the 
CCD-gaps in the two sets of instruments.  The areas of the CCD gaps are 
excluded from the following analysis.

\subsection{X-ray maps}

To better visualise the X-ray emission from \scg\ we have smoothed the
data with an adaptive smoothing algorithm.  We show the isointensity
contours derived from the adaptively smoothed EPIC-MOS data in the
0.3-2.0 keV energy range superposed onto a V-band image in
Fig.~\ref{figure1}, together with the smoothed X-ray map with the
contours indicating the optical sources in the field.

It is evident that most of the emission arises from individual 
sources, but some low surface brightness diffuse emission 
is also present between the galaxies.

All members of SCG0018-4854 are detected: NGC~87, NGC~89, 
NGC~92 are clearly visible in Fig.~\ref{figure1}. A source coincident with 
NGC~88 is also detected, although it is 
less clearly visible due to a very close brighter source to the W, 
coincident with a faint optical source, most likely a background
object. ESO 194-G 013 also coincides with a source 
in the EPIC-pn and EPIC-MOS2, very close to the edge of the
field (not covered by Fig.~\ref{figure1}). 
The source to the E of NGC~92 is most likely another background object,
barely visible on the UK Schmidt Telescope
data.  

The space between the galaxies appears to be filled with genuine 
emission,  as will be quantified later, which we interpret as due
to a hot intergalactic medium (IGM). 

\subsection{The extended diffuse component}

To quantify the amount and distribution of this diffuse X-ray
component, the determination of the background shape and level is
important, in particular considering that we want to measure a
very low surface brightness emission.  Two methods of determining
the background were tested, using the ``blank sky'' data available
online vs. using a local background from our data. After careful 
consideration (see below), we resolve to use a local determination 
of the background, while using some information from the blank sky
field to make some required adjustments to the surface brightness
levels.

The medium filter event files for the ``blank sky'' data available 
online\footnote{at
http://xmm.vilspa.esa.es/external/xmm\_sw\_cal/back-
ground/blank\_sky.shtml or kindly provided by A. Read and J.  Carter,
that constitute and improved version of the on line data with a
smaller residual at the centre} \citep[see][ for a comprehensive 
discussion on the characteristics and preparation of the files]{carterread}
have been cleaned from the high background episodes in a
manner consistent with the \scg\ data and the files have been rotated
to the same angle as our observation with the $skycast$\footnote{see
http://www.sr.bham.ac.uk/xmm2/scripts.html} program.

\begin{figure}
\psfig{figure=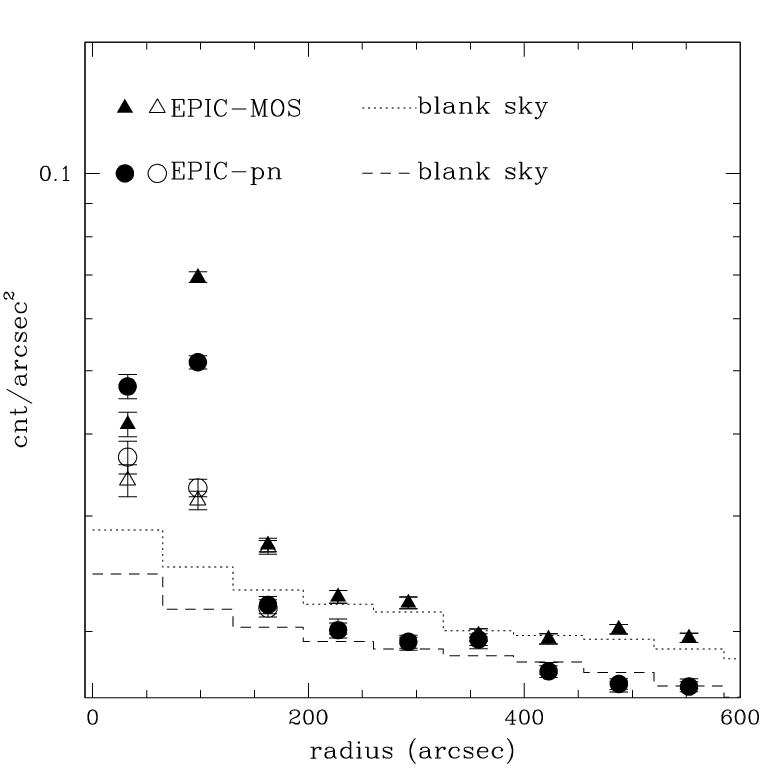,width=8cm}
\caption{Radial profile of the total emission in \scg\ in the 0.3-2.0 keV
band, from EPIC-MOS and EPIC-pn separately, centred between the 4
member galaxies. Detected sources are included in the filled symbols, and 
excluded in the open ones (shown only for the innermost regions).
The background from the blank sky
fields is also shown, renormalised with the exposure time ratios
(adjusted with a 
correction of $\sim$9\% in the EPIC-pn).
}
\label{figure2}
\end{figure}

\begin{figure}
\psfig{figure=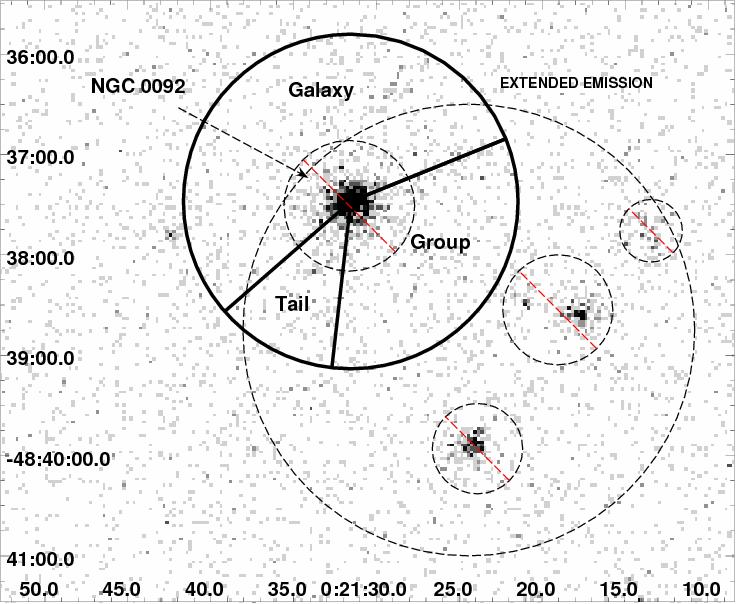,width=8cm}
\caption{Regions used to derive the spatial and spectral parameters of
the different sources discussed in the text (see Figs~\ref{figure3},
~\ref{tail}) plotted on the X-ray image of \scg.  The large circle delimited
by dashed lines indicates the region used to evaluate the
characteristics of the extended emission. The smaller circles with the
diameters indicated in red  are masked out to exclude
individual sources.  The regions delimited by thick black lines centred
on NGC 92
indicate the angular sectors used in Figure~\ref{tail} to estimate the
excess due to the tail in NGC 92.}
\label{regions}
\end{figure}

Fig.~\ref{figure2}\ shows the comparison between the azimuthally
averaged radial profile obtained from our observation of \scg\ and 
from the blank sky data, normalised to the \scg\ profile at large 
radii (note that the normalisation is consistent within $\le$ 10\% 
with the ratio of the exposure times). 
We show both the full profile, inclusive of all sources (with filled
symbols), and the profile of the ``residual" emission, where all detected
sources have been masked out, with the regions shown in
Fig.~\ref{regions}. 
We use the 0.3-2.0 keV band to maximise the relative contribution
to the total emission of the soft component, while retaining a 
wide-enough band to include also the harder components. The centre 
of the profile is at 00$^h$21$^m 24\fs25$, -48\degr38$'51\farcs7$, 
in the central area between the 4 galaxies of the group. 
The good agreement between 
the data and the blank sky profile shapes at radii larger than 
$\sim 200''$ suggests that these could be used to estimate the 
background in the central area of the group, and indicate a 
significant excess emission at least out to $\sim 150''$.

\begin{figure}
\psfig{figure=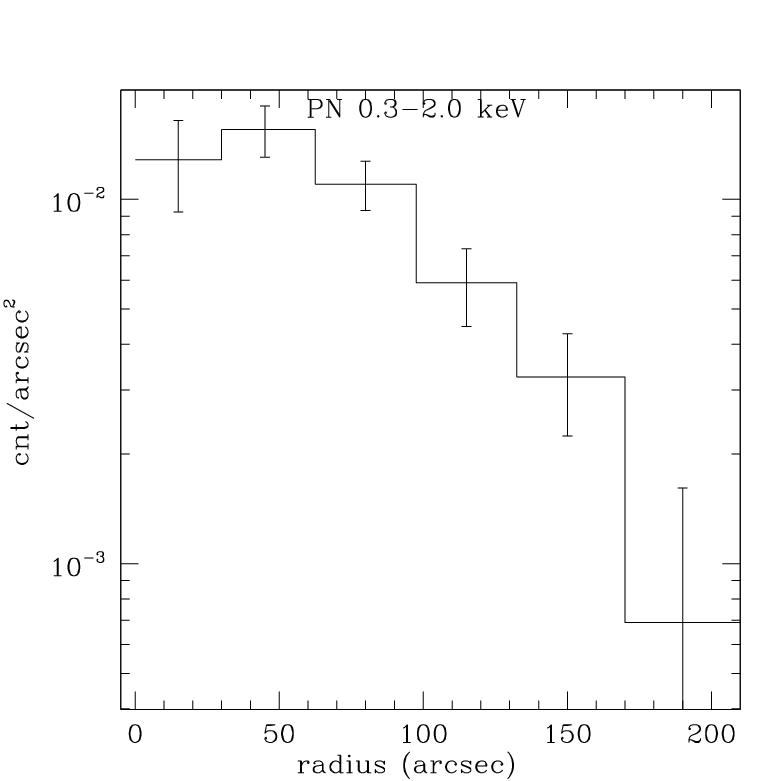,width=8cm}
\caption{ Net profile of the group component (EPIC-pn, 0.3-2.0 keV)
excluding all detected sources, centred midway between the galaxies.
}
\label{figure3}
\end{figure}

\begin{figure*}
\resizebox{18cm}{!}{
\psfig{figure=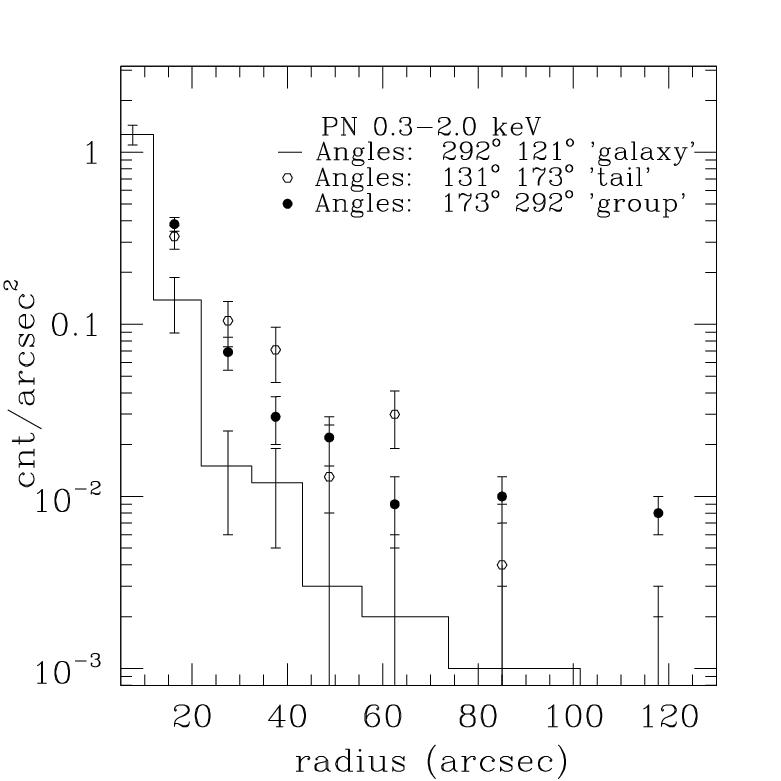,width=18cm}
\psfig{figure=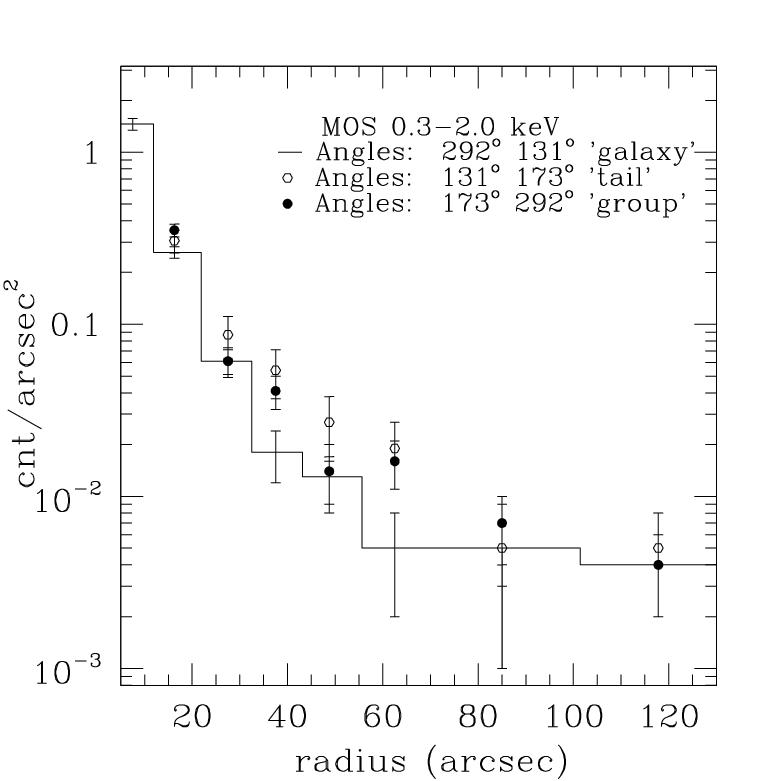,width=18cm}
}
\caption{LEFT: EPIC-pn and RIGHT: EPIC-MOS radial distribution of the
emission in the stated angular sectors, measured counterclockwise from
N.  A graphical indication of the azimuthal sectors used can be found
in Figure~\ref{regions}.  }
\label{tail}
\end{figure*}

However, an analysis performed over narrower energy bins indicates that 
the normalisation of the two sets is different at different energies, 
requiring a correction of up to $\sim$ 17\% to raise the background 
to the observed value at energies below 2 keV, and $\sim$ 6\% to 
lower it for energies above 2 keV, relative to the average value 
for the 0.5-2.0 keV band. This is most likely a consequence of the 
difference in the spectral shape of the blank sky relative to our 
field background that converts into slightly different normalisations 
in different bands \citep[see also][and references therein]{carterread}.  

Since it appears that outside a radius of $\sim 200''-250''$ the 
spatial shape of the emission is entirely consistent with the rescaled blank
fields, and that the shapes of the blank sky profiles are  virtually
the same at all energies below $\sim 5$ keV, we conclude that a local
determination of the background  is possible, provided the blank sky 
spatial profile is used to rescale it to the appropriate level at the source 
area, which is what we have done.

The net profile from the diffuse emission in \scg\ is shown in
Fig.~\ref{figure3}, where all detected discrete sources have been
excluded.
Emission is clearly visible out to a radius of $\sim 2'$, with a 
total of $\sim 410\pm40$ net EPIC-pn counts.  Since
all detected sources have been masked out, we expect that the excess
is truly from emission other than from the individual galaxies:
photons in the wings of the Point Spread Function (PSF)
due to the sources detected
individually can account only for a fraction of the diffuse
emission.  Given that the masking regions (circles of r$\sim
18''-40''$, depending on the source strength and the presence of more
than one source in a small region, see Fig.~\ref{regions}) contain
$\sim$ 70-80\% of the emission of the point sources (see the
XMM-SOC-CAL-TN-0029 and XMM-SOC-CAL-TN-0022 documents online at {\tt
http://xmm.vilspa.esa.es}\footnote{\citet{Bel03} have shown that
extended sources like galaxies can be treated as point sources with
little loss of accuracy for this purpose}), we estimate that 
$\sol$ 20\% of the counts in the extended component could be contributed from 
the individual sources detected. 

\begin{figure*}
\resizebox{18cm}{!}{
\psfig{figure=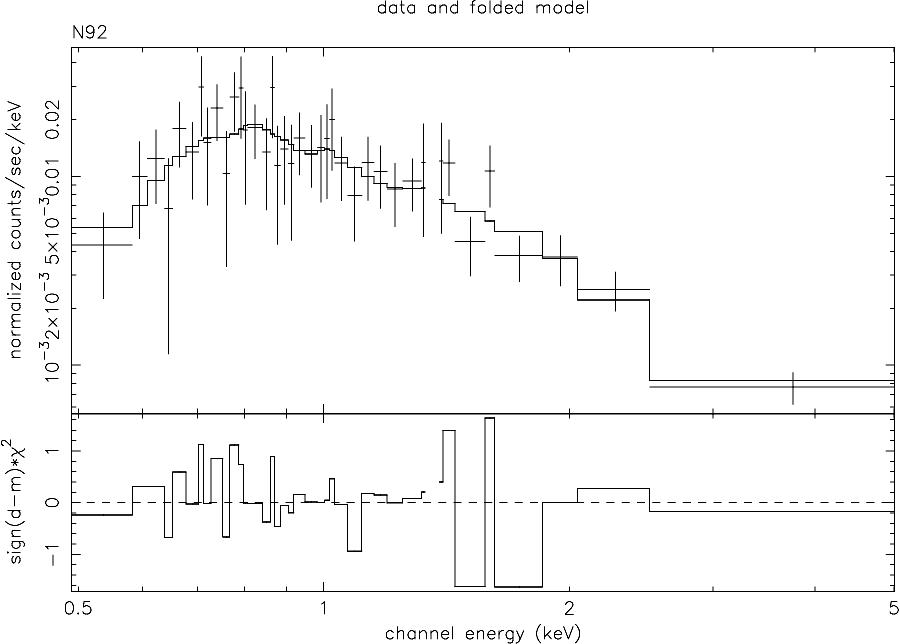,width=17.5cm}
\psfig{figure=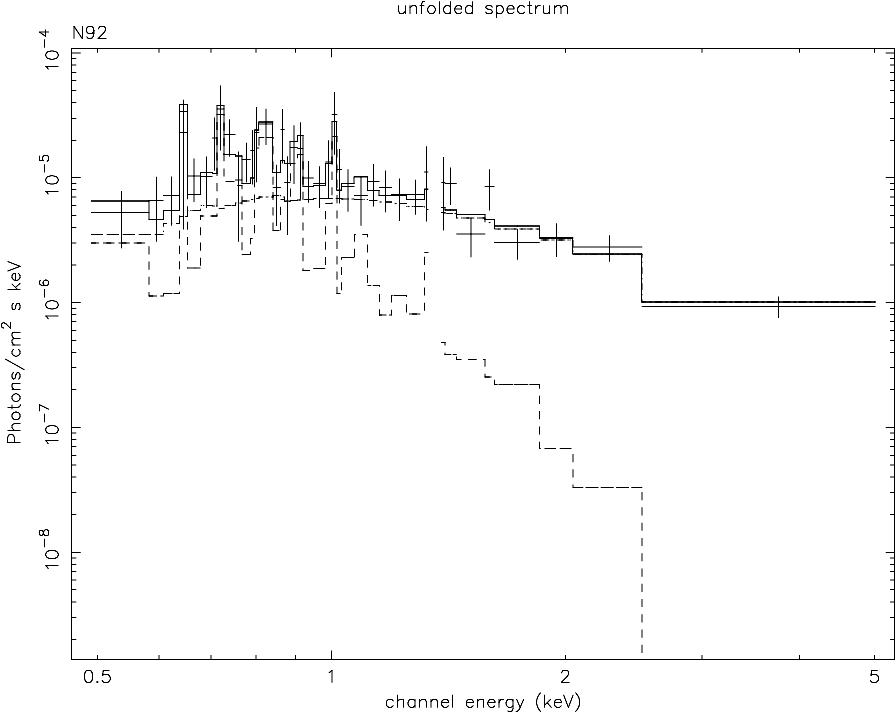,width=15.5cm}
}
\caption{EPIC-pn spectral data (crosses with error bars) with the best fit model (histogram)
and the distribution of the significance ($\chi^2$) as a function of energy (left panel)
and the unfolded spectrum with the individual spectral components  (right panel) for NGC 92. 
 }
\label{spec-92}
\end{figure*}

\subsection{The tail of NGC~92}
A rather unexpected feature can be observed in the raw data as an
excess to the SSE of this galaxy.  The statistical significance of 
the excess emission is limited by the poor photon statistics and 
cannot be seen as a separate feature in the iso-intensity contours 
of Fig.~\ref{figure1}, possibly because of
its closeness to the relatively bright source associated with the
galaxy.  However, the contours centred on the galaxy are clearly
elongated in this direction.  The location of the feature is
intriguing, since it coincides with the base of the tail to the
South seen prominently in H$\alpha$ and \hi\ \citep{Temporin, Pompei07}.  
Again,
we make use of a radial profile plot (Fig.~\ref{tail}) to compare the
distribution along the SSE direction and the expected distribution
from the same source in the N-NE directions (see Fig.~\ref{regions}).
There is a clear excess of emission in the SSE direction (tail)
relative to the northern half (the galaxy only).  This latter
component can be parameterised by a combination of a central point
source and a $\beta$-type model $\Sigma_x \approx (1+{r \over
r_0}^{-3\beta+0.5}$), with r$_0 \sim 9''$ and $\beta \sim
1.3$, consistent with the contribution from a nuclear source and gaseous
emission from the ISM (see next section). In the tail sector instead
the parametrisation requires a significantly flatter $\beta$-type
model with $\beta \sim 0.53\pm 0.1$ (1-$\sigma$ confidence), 
indicative of a shallower radial
gradient.  Although the excess in this region could in part be due
to the emission from the gas in the group (see above), there appears
to be an excess even above this component, as shown by the
comparison with the profile towards the group centre (Fig.~\ref{tail})
outside of $\sim 20''$ radius.  We can measure an excess of $\sim
35\pm8 , 45\pm10 $ counts in the $\sim 20''- 80''$ region in the
EPIC-pn and EPIC-MOS data, respectively, relative to the model of
the emission from the galaxy in the northern azimuthal sector.  Assuming a
plasma with kT$\sim 0.3$ keV (the average value in the whole galaxy,
see next section), the tail could contribute L$_x \sim 4 \times
10^{39}$ \ergsec\ (0.5-2.0 keV) to the total X-ray luminosity of
NGC~92.

\subsection{Spectral properties of the emission}
To measure the spectral properties of the emission from different
sources we have extracted photons in different size regions.  We have
corrected for the areas lost due to CCD gaps in the
EPIC-pn.  We use the EPIC-MOS data when statistically significant to
confirm the EPIC-pn results, or to assess the total luminosities, when
the EPIC-pn gaps cover a significant fraction of the source area.

\begin{figure}
\psfig{figure=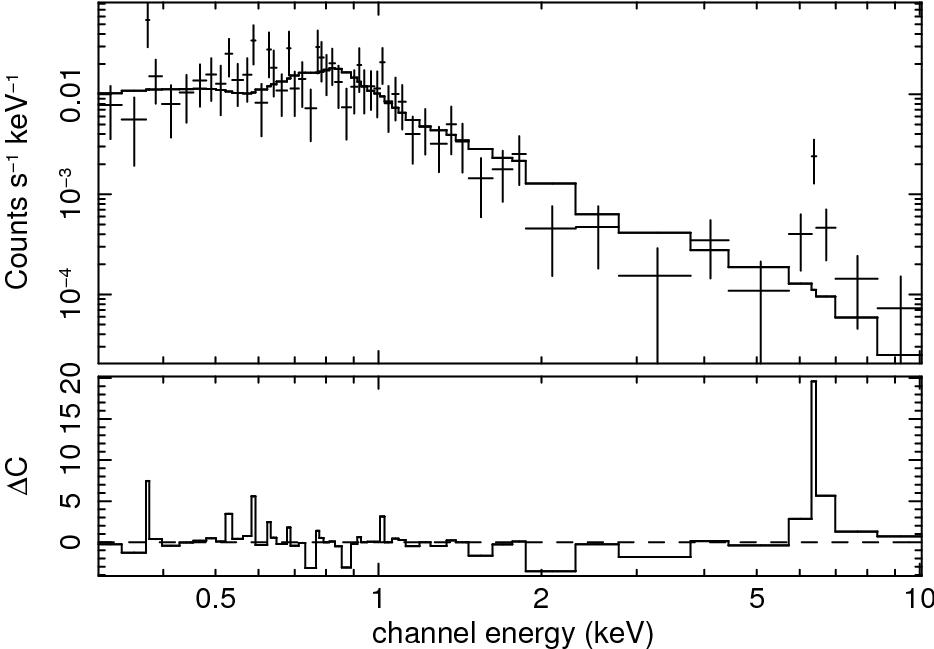,width=8.5cm}
\caption{EPIC-pn spectral data for NGC 89.  The data are binned at $\sim 2 \sigma$ in net counts per bin at low energies and at lower significance, 
but avoiding a ``negative" signal, at high energies above 2\,keV. 
The model shown and the $\Delta$C distribution refer
to the mekal + power law model (see
text).}
\label{spec-89}
\end{figure}

The background is in all cases taken from an annulus outside the group 
area, rescaled as discussed in the previous section.  We have also
considered the region between galaxies as background for the sources
associated with the group members, and we found entirely consistent
results in the derived quantities, with only a small decrease in the net 
count rates ($4-20$\%, for NGC 92 and NGC 89 respectively). 
We binned the data to increase the photon statistics per
spectral bin to ensure that the $\chi^2$ statistics can be applied.  
We have used a binning based on a minimum number of counts in the 
total data (typically 30 counts per bin) or
based on a minimum signal-to-noise ratio 
in the net 
data, depending on the observed number of photons, while at the same
time ensuring that there was a reasonable number of bins for the
spectral fit.  The data are fitted in XSPEC (version 11.3.1) with 
either a
plasma with low energy absorption (e.g. MEKAL) or a combination of
plasma and bremsstrahlung or power law to account for high energy
tails.  We adopt the tables of \citet{Wilms} for the abundances and a
Galactic value of N$_H = 2.7 \times 10^{20}$ cm$^{-2}$ \citep{dickey}. 
Quoted errors correspond to a $\Delta \chi^2$=2.7 unless specified otherwise. 

\begin{figure*}
\resizebox{18cm}{!}{
\psfig{figure=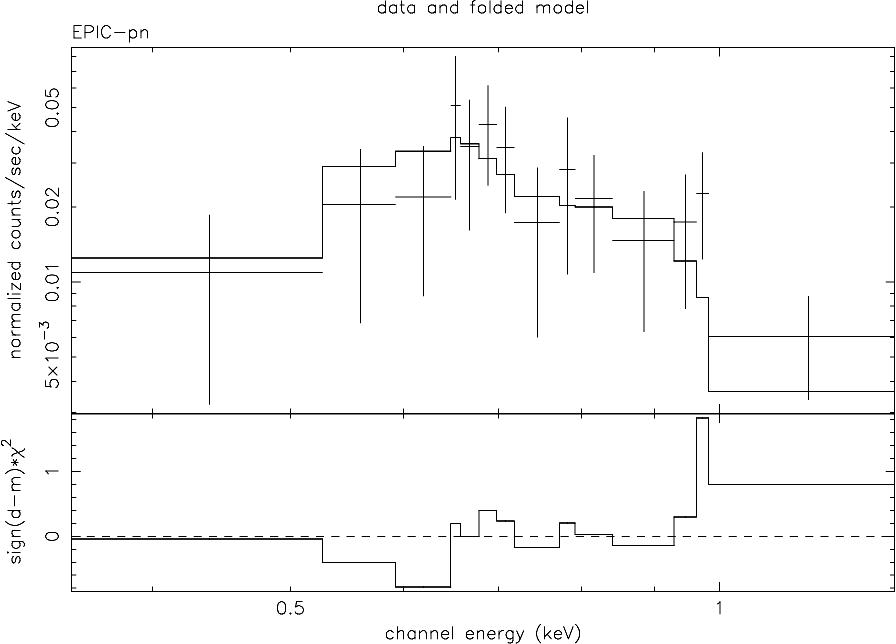,width=17.5cm}
\psfig{figure=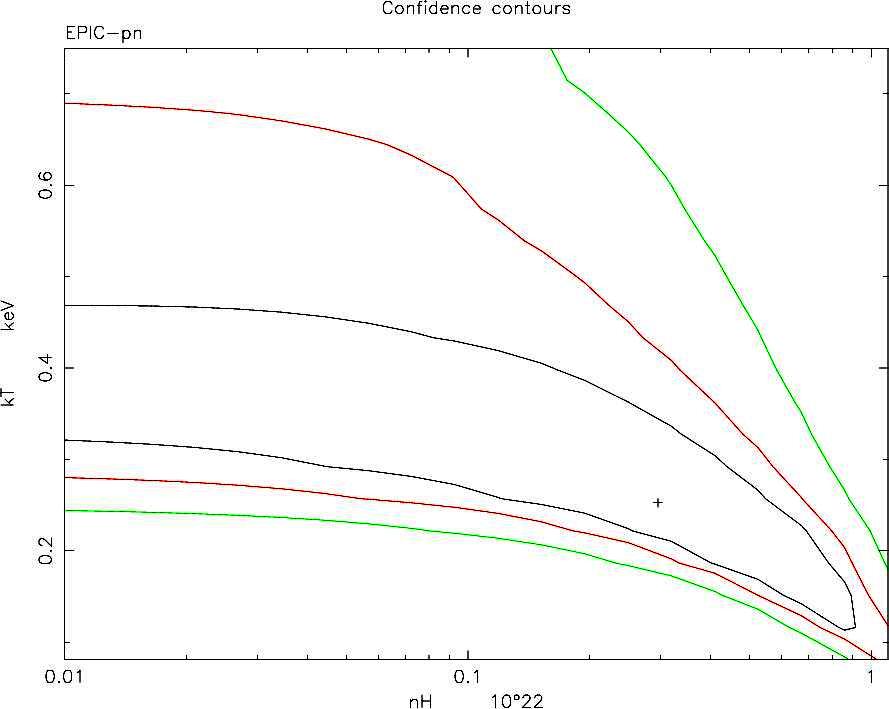,width=15.5cm}
}
\caption{EPIC-pn spectral data and best fit model (crosses with error bars and histogram
respectively) in the left panel, and  confidence contour regions
(at 68, 90 99\% level) for N$_H$ and kT values (right
panel) for the diffuse 
extended emission.  }
\label{spec-ext}
\end{figure*}

\noindent\underline{NGC~92}
The  X-ray spectrum of NGC~92 can be parameterised by a combination 
of a power law with $\Gamma \sim 2.2 \pm 0.6$ and a solar abundance plasma with 
kT$\sim 0.3\ [0.22-0.55]$ keV (Fig~\ref{spec-92}). 
A relatively large amount of low 
energy absorption is required, corresponding to N$_H \sim 4 \times 10^{21}$ 
cm$^{-2}$, consistent with the presence and amount of neutral hydrogen 
observed in HI emission \citep[visible in Figure 5
in][]{Pompei07}.  The intrinsic luminosity of the power law component,
L$_x  \sim 4 \times 10^{40}$ \ergsec\ in the 0.5-10 keV band, 
points to the presence of
an AGN, albeit of very low luminosity.

The absorption corrected 
plasma luminosity of $\sim 2 \times 10^{40}$ \ergsec (0.5-2 keV, 
within a 30$''$ radius), is relatively high for a normal spiral galaxy,
but is consistent with the presence of
starburst activity. 
Unfortunately we cannot study the spectral properties of the emission
in different regions; however, given its extended nature, also
testified by the fit parameters with a $\beta$-model (see previous
sections), we speculate that the emission is not to be attributed
exclusively to the nuclear starburst region, but is most likely
distributed throughout the body of the galaxy.

\noindent\underline{NGC~89} 
The energy distribution of the signal allows a $\sim 2 \sigma$ binning of the net spectrum only
at low energies, up to $\sim 2$ keV.
In the 2-10 keV range, too few photons ($ 28 \pm 8$) are detected, for a meaningful spectral
analysis. However, taking into account that this object has a Seyfert 2 nucleus, from the
optical line ratios, we will first use the spectral data at low energies, to obtain information
on the X-ray spectral properties of the object.  We will then add the high energy range
to obtain some information on possible nuclear emission. 

A single component plasma model does not fit the data well; a single
temperature bremsstrahlung or a power law model could give
significantly better fits, but with unreasonable parameters (kT$\sim
0.5$ keV, which is also obtained with a zero abundance MEKAL model, or
$\Gamma=4$, respectively).  We obtain a reasonable fit ($\chi^2 \sol
1$ for 16-17 degrees of freedom in EPIC-pn) with a two-component model
(two thin plasmas at 0.25, 1.4 keV, 50-100\% solar abundance or one
plasma at $\sim 0.4$ keV, 100\% solar abundance and a power law
component with $\Gamma$ fixed at 1.9).  

Either models could be interpreted in the framework of the emission from
spiral and starburst galaxies, resulting from to the combined contribution
of a population of
unresolved compact sources and a diffuse emission, most likely
due to a hot phase of the ISM.
The line-of-sight absorption is
always consistent with the Galactic value.  The unabsorbed total flux
from this object is f$_x \sim 1.4 \times 10^{-14}$ \ergcmsec, which
corresponds to a total intrinsic luminosity of L$_x \sim 4\times10^{39} $
\ergsec (0.5-2.0 keV), with a contribution of 1/3 from the softer 
(plasma) component and 2/3 arising from the harder (higher 
temperature/power law) component, respectively.

The power law component could however also be interpreted as due to a central AGN, visible in
the optical data. However, its intrinsic luminosity would only be
L$_x \sim 4\times10^{39} $\ergsec (2.0-10 keV), indicative of a very faint AGN, unless it were
interpreted as the scattered component of a heavily obscured object. Optical line emission
ratios
indicate a Seyfert 2 nucleus in NGC 89, so the presence of an absorbed central source is quite
plausible. This interpretation is
supported by the evidence of an excess at the expected position of the Fe line. Given the
low statistics in the high energy bins, we have neglected to take into account the significance
of individual bins, and simply rebinned the data  in such a way that no bin contains a
``negative" signal.  The resulting spectrum, shown in Fig.\ref{spec-89}, is then compared with an
absorbed AGN model, in addition to a plasma component to account for the low energy emission. 
To account for the small number of counts in each bin we have used the C statistic
\citep{cash}.
An excess of $\sim 10$
counts is evident at the Fe line energies.  If we assume a typical absorbed AGN spectrum, the
data are consistent with an intrinsic luminosity of a few $\times10^{41} $\ergsec, and an
absorption greater than a few $\times10^{24} $ cm$^{-2}$.  The excess at $\sim 6.5$ keV is 
 at a $>99$\% significance from the C statistics
($\Delta$C=20), and it is consistent with the expected narrow 6.4~keV Fe line emission seen in local
Seyfert
galaxies \citep[e.g. see][]{Nandra}.

\noindent\underline{Extended intra-group emission}
EPIC-pn data have been extracted from a region of $\sim 2\farcm25$,
excluding the CCD gaps and all detected discrete sources (the 4 
galaxies and the source W of NGC ~88, see Fig.~\ref{regions}).  
The background is rescaled by $\sim 10$\%
to take into account the effect of vignetting, as explained in
previous sections.  The data are unfortunately of low signal-to-noise,
so the results are highly uncertain, which reflects also on the
derived quantities (Fig~\ref{spec-ext}).  If we assume a plasma model
(MEKAL) with abundances fixed at 30\% cosmic, we derive a temperature
of $\sim$0.2 ([0.1-0.5] keV, see
Fig.~\ref{spec-ext}, right panel) and a higher than Galactic 
line-of-sight absorption,  N$_H \sim 4 \times 10^{21}$ cm$^{-2}$. This converts
into an emitted flux f$_x \sim 1 \times 10^{-13}$ \ergcmsec, in the
0.5-2.0 keV band, strongly dependent on the exact values of N$_H $ and
kT assumed.  If we take into account the area lost due to the CCD gaps
and the sources excluded from the analysis, the extended component
flux should be raised by $\sim 40$\%, assuming a flat surface
brightness distribution in the area.  Therefore the total intrinsic luminosity
of this component is of the order of $8.5 \times 10^{40}$ \ergsec\ for the
stated best fit parameters.

\noindent\underline{Point source West of NGC~88}. 
The data can be fitted with a power law with $\Gamma = 1.8$ [1.4-2.0]
with mild absorption N$_H \sim 8 \times 10^{20}$ cm$^{-2}$, for an
unabsorbed flux of $1.0 - 1.6 \times 10^{-14}$ \ergcmsec in the
0.5-2.0 and 2.0-10.0 keV range respectively. This would imply a
luminosity of $\sim 8 \times 10^{39}$ \ergsec (0.5-10.0 keV), if the
source were at the distance of the group.  While this would be
consistent with an ultra luminous source in NGC~88, its coincidence with a faint
optical counterpart makes it more likely to be a background source.

\section{The X-ray picture}

The sensitivity achieved with the XMM-Newton observation has provided firm
evidence of a hot intergalactic medium in which galaxies are embedded in
the spiral-only, dynamically young system \scg, analogous to what has
been known to permeate more evolved systems.  With these data we have
detected and characterised the emission from the member galaxies (most
clearly from a, b, and c, or NGC 92, NGC 89 and NGC 87 respectively) and
from a diffuse component, and have investigated the spatial distribution
and the spectral features for the galaxies and for the intergalactic
medium in some detail.

\subsection{Individual galaxies} 

All galaxies of the group are detected, at a level
comparable to other systems of the same optical characteristics, both in the field and 
 in groups 
\cite[see][]{Helsdon01}.  
When measured, the spectral properties are consistent with a
contribution from a hot thin plasma and unresolved X-ray binaries,
typical of the emission of late type galaxies with elevated star
formation activity.

For NGC~87 and NGC~88 and  ESO 194-G 013, not detected with enough counts for a proper
spectral analysis, we derive a luminosity L$_x \sol 10^{39}$ \ergsec\
from a simple rescaling from net counts into fluxes and
luminosities, assuming a power law model with $\Gamma=1.7$. 

NGC~89 has a higher luminosity, L$_x \sim 4 \times 10^{39}$ \ergsec,
in the 0.5-2.0 keV band, consistent with the combination of a hot ISM at $\sim$0.4 keV and
the unresolved contribution of binary systems. In spite of the very low statistics at higher
energies, the data are consistent with the presence of a heavily absorbed (N$_H > 10^{24}$
cm$^{-2}$) 
AGN, of intrinsic L$_x \sim \times 10^{41}$ \ergsec (2-10 keV). The evidence for such 
a component is given by
an excess of counts at the expected position of a Fe line.  The equivalent width of the 
Fe line would be of several keV (poorly measured) against the very weak hard X-ray 
continuum that we can infer for NGC 89, but would be a few tens of eV against an AGN with an
unabsorbed luminosity of a few $ \times 10^{41}$ \ergsec, i.e. consistent with
the narrow iron lines seen in local Seyferts \citep[i.e.][]{Nandra}. Moreover
the energy of the line is formally consistent with 6.4 keV
(6.46$\pm$0.08 keV), as would be expected from neutral Compton reflection. 
This is in excellent agreement 
with the optical evidence of a Seyfert 2 nucleus, although the maximum luminosity allowed by
the data, for a reasonable set of parameters for the AGN spectrum, is significantly lower than 
predicted for example on the basis of the observed [OIII] line
\citep[see, e.g., the NGC 1068 example by ][]{colbert},  
but in line with the scatter observed for  local AGNs \citep[see][]{heckman,panessa}. 
Interestingly, this is
the only member of \scg\ without trace of \hi\ gas \citep{Pompei07}.

NGC~92 is the brightest of the galaxies, with a total luminosity L$_x
\sim 2 \times 10^{40}$ erg s$^{-1}$, in a hot plasma component.
A mildly absorbed, low luminosity (L$_{(0.5-10 keV)}
\sim 4 \times 10^{40}$ erg s$^{-1}$) nuclear source is also inferred from the
morphology and spectrum. 
A population of binary sources, which must also be present, cannot be
detected separately from the central ``AGN-like" component. 
A central radio continuum source is also detected in this galaxy
\citep[P$_{1.34 
GHz} \sim 4 \times 10^{22}$ W Hz$^{-1}$,][]{Pompei07}, which so far has been 
associated with starburst activity.  At the present time there is no
measure of the compactness of the source nor good spectral index information, to
help understand whether the radio source could be associated with the possible
central AGN. 

For both galaxies, the luminosity measured in the ``plasma"
component is higher than what is predicted by the emissivity of the old
stellar population, and should therefore most likely be a truly hot
ISM component \cite[see][]{revnivtsev}.

The global star formation rate measured for NGC 92 is 14.7 M$_\odot$
yr$^{-1}$ \citep{Pompei07}, which is significantly higher than that of
other starforming galaxies like M82, NGC1482 or NGC3079 \citep[$<10$
M$_\odot$ yr$^{-1}$][]{Strickland}, although the total luminosity in
the diffuse X-ray emission is lower. 
If we use
the spectral results and attribute the emission modelled with a plasma
spectrum to a hot ISM, we can derive an average gas density of n$_e
\sim 7.5 \times 10^{-3}$ cm$^{-3}$ and a total mass of M$_{gas} \sim 3
\times 10^8$ M$_\odot$ in the hot phase, under a simple assumption of
a spherical distribution within r$\sim 7.5$ kpc. This is $\sim
10\times$ lower than what is detected in the neutral cold phase
\citep[\hi][]{Pompei07}, but relatively high with respect to the hot
ISM detected in the disk and halos of star-forming late type galaxies
\citep{Tuellmann} and comparable to the amount of hot gas detected in
HCG80a \citep{Ota04}.

A peculiar extension to the SE of NGC~92 coincident with a tidal
feature observed in \hi\ and H$\alpha$ is also detected.  While
current data do not allow us to properly separate it from the emission
of the galaxy, its existence is clearly shown in the elongation of the
isointensity contours in the soft energy map and in the significantly
flatter $\beta$-model required to fit the radial profile in the SSE
direction.  As will be discussed later, the formation of the X-ray tail
is unlikely of stripping origin.  Most likely, the tail is the result of
tidal forces acting
on NGC92 have been efficient in
dislocating the \hi\ gas;  in situ star formation, which produces the
spectacular H\ii\ regions and H$\alpha$ emission observed in the tail, can also
produce X-ray emitting sources (gas and/or binaries).

\subsection{The hot IGM}

\scg\ has a very tight configuration and optical studies already
support the hypothesis that it is a truly bound system. In fact the
four galaxies of the group occupy an extremely small region in space,
all being located within a circle on the sky of $\sim$ 25 kpc.
Assuming that this dimension is indicative of the true physical radius
of the group, the density within the group obtained is a record one
also for such tight configuration systems, at log$({gal \over
{Mpc^{-3}}}) = 4.88$, \citep[see][]{Pompei07}.  Also the
line of sight velocity dispersion of the group is very low, amounting
to only 120 km s$^{-1}$, although the number is highly uncertain due
to the paucity of the velocity space sampled [but typical of these
groups].  As a further indication of the high density of this system
and of the high level of interaction among members, all four group
members have a disturbed optical morphology, and a very peculiar \hi\
distribution mostly dislocated with respect to the bulk of the stars
\citep[NGC 89 is in fact completely stripped of gas,][]{Pompei07}.

With these XMM-Newton observations we have detected extended X-ray
emission at the centre of the group, in a region  of 
$\sim$35 kpc radius.  We cannot study the morphology of such
component in detail, due to the very limited statistics of the
detection; however, we can exclude that the emission is the result of
the combined emission from the individual members, 
and we claim it is a genuine detection of a hot IGM
at a level of $\sol10^{41}$ \ergsec.

To characterise the gas, we have assumed ionisation equilibrium and
30\% cosmic abundances, and we derive a relatively low temperature of
kT $\sim 0.2$ keV.  This is the usual approximation, in spite of the fact that 
the system is most likely not in a stable configuration, also given
the optical evidence of interactions among group members. However,
with current statistics, it would be unrealistic to attempt more
complex modelling of the data.  This result makes this group the
coolest detected so far, even among the few spiral only groups studied
in the X-ray band.

In this model, assuming that the emission is distributed in a sphere
with radius $135''$ (33 kpc), and correcting for the area lost due to
the sources excluded from the analysis, the mean density and total gas
mass are n$_e \sim $5$\times 10^{-3}$ cm$^{-3}$ and M$_{gas} \sim$ 2 $\times
10^{10}$ M$_{\sun}$ respectively, estimated from the emission measure
obtained from the spectral results.  At the centre, within 20 kpc, where
the radial profile is consistent with a constant distribution, the gas
density increases to 0.01 cm$^{-3}$ with a cooling time $\sim 2\times
10^{9}$ yr.

The total amount of hot material is significantly lower than observed
in more typical groups \citep[e.g.][]{M00}, but comparable to the gas
masses derived for low velocity dispersion groups by \citet[][a few
$\times 10^{10-11}$ M$_{\sun}$ within comparable radii]{Helsdon05}.
The inferred temperature is also significantly lower than those measured
in more evolved groups (1-2 keV); unfortunately at these low temperatures
the L$_x-\sigma$ and L$_x-$T relations for groups and clusters have a
significant scatter, so consistency with such relation is of little
significance; however, it appears that this group has a higher than
expected X-ray luminosity for its measured temperature and for its $\sigma$
value relative to the extrapolation from the cluster relation \citep[see
][]{Helsdon05}.

Given the low velocity dispersion of the group \citep[$\sim120$ km
s$^{-1}$][]{Pompei07}, the gas density calculated is not enough to
explain the formation of the HI tail in N92 with ram pressure stripping. Using
typical values for the cold ISM content in a galaxy 
and $P_{\rm ram} = \rho_{\rm ICM} \, v_{\rm gal}^2 =
\Sigma_{\rm gas} \frac{v_{\rm rot}^2}{R_{\rm strip}}$ \citep[where
$\rho_{\rm ICM}$, $v_{\rm gal}$, $\Sigma_{\rm gas}$ and $R_{\rm
strip}$ are the mean ICM density, the velocity of the galaxy relative
to the ICM, the surface density of the galaxy's ISM, and the minimum
stripping radius of the galaxy,
respectively, see][]
{Trinchieri2007}, the observed quantities for $\rho_{\rm
ICM}$ and v$_{\rm gal}$ imply $R_{\rm strip}=9 \times 10^{23}$ cm,
$\sim 300$ kpc, well outside the galaxy boundaries.  For stripping to
be efficient, the transverse velocity should be an unlikely
factor at least 6-7
larger than the observed velocity dispersion of the group.

\subsection{\scg\ in context}

Very few dynamically young, spiral dominated systems like \scg\ have
been observed in X rays and even fewer have a measured IGM.  No IGM
was detected in the groups with a spiral fraction of $\sim 1$ in the
\citet{Mul03} atlas, while only a few in the groups studied as part of
the GEMS project \citep{OsmondPonman2004} have been classified as
having a group-scale diffuse emission (e.g. $>60$ kpc in radius), both
based on lower resolution ROSAT data.  In HCG 16, one of the first such groups well
studied and much discussed in the literature, gas was 
unambiguously detected only recently with XMM-Newton \citep{Bel03}, at a
level of $\sim 10^{41}$ \ergsec.

HCG~16 is in many respects very similar to \scg: it is composed of 4
luminous late type galaxies in a small area \citep[median projected
galaxy-galaxy separation of $\sim 60$ kpc  h$_{70}^{-1}$][]{Hickson92},
with evident signs of activity in their nuclei, and a velocity
dispersion of $\sim 100$ km s$^{-1}$.  The X-ray properties appear
also very similar: all 4 galaxies are detected as individual sources
(both due to the AGN and starburst activity) already by the ROSAT
satellite \citep{SaraccoCiliegi, Ponman, dossantos}, while the diffuse
emission recently characterised with XMM-Newton has a comparable
luminosity but slightly higher best fit temperature \cite[$\sim 0.5$ keV,
although within the uncertainties; note that the ROSAT data suggested a
cooler temperature of 0.27 keV][]{dossantos} than \scg.  
Only two other systems,
with a spiral fraction of $\sim 1$, have been studied with Chandra or
XMM-Newton: HCG 80 \citep{Ota04} was not detected by Chandra, with an
estimated limit to the luminosity of the diffuse component at L$_x
\sim 6 \times 10^{40} h^{-2}_{70}$ \ergsec, in spite of the larger
velocity dispersion ($\sigma_v \sim 300$ km s$^{-1}$) of the system;
low level diffuse emission is measured in recent XMM-Newton images of
the Cartwheel, again at similar luminosities and gas temperatures
(Crivellari et al., in preparation).

Our results seem to suggest that, 
to obtain a more significant hot IGM component, we need groups with a
somewhat lower spiral fraction and/or in a more evolved stage of
evolution, so that they have had time to accumulate a more
significant amount of hot IGM and/or the interaction among members has
already been able to heat it to higher temperatures. Some of
the famous examples include HCG92-Stephan's Quintet
\citep{Trinchieri2003,Trinchieri2005}; HCG
90
\citep{hcg90ref}; and an earlier ASCA detection of diffuse
emission in HCG 57 \citep{Fukazawa}, all characterised
by a hotter IGM (kT $\sim 0.7-1$ keV), a spiral fraction of
$\sim0.5-0.7$,
relatively complex morphologies 
and somewhat higher luminosities.

\section{Conclusions}

The detection of diffuse emission from \scg\ indicates that this system contains 
a hot intergalactic gas with a mass of $\sim$1\% the mass measured in galaxies.
The amount and temperature of the
gas are about one order of magnitude lower than in more evolved
groups, where the gas is usually peaked onto the brightest
(early-type) member, at temperatures of 1-2 keV, and has a significantly
larger extent.  The temperature is broadly consistent with what is expected from gas in equilibrium 
in the potential defined by the measured velocity dispersion
of the group.  Although at lower levels than in other systems, the evidence of
such a component in this group is important because it can be used as
additional evidence that a) the galaxies are physically related in a
bound system, and b) the IGM can be a source of disturbance and
participate in the production of distortions/tails observed in the
galaxies, even though it alone cannot explain the production of
e.g. the tail of NGC~92.  It is worth mentioning that the
characteristics of the hot gas detected in \scg\ are comparable to the
values derived for our own Local Group from indirect measures
(absorption features in the spectra of background sources): T$\sim 2-5
\times 10^6$ K, N$_e < 2 \times 10^{-4}$ cm$^{-3}$, and a scale of
$\sim 140$ kpc \citep{rasmussen}.  

A larger sample of objects however still needs to be sampled at the
proper sensitivity to provide the missing evidence of whether these
are special cases or all spiral dominated systems have a hot, though
cool and weak, IGM.

\begin{acknowledgements}
This research has made use of SAOImage DS9, developed by Smithsonian
Astrophysical Observatory and of the NASA/IPAC Extragalactic Database
(NED) which is operated by the Jet Propulsion Laboratory, California
Institute of Technology, under contract with the National Aeronautics
and Space Administration.  Both Ciao and XMM-SAS softwares have been
used to reduce the data. 
This publication makes use of data products from the Two Micron All
Sky Survey, which is a joint project of the University of
Massachusetts and the Infrared Processing and Analysis
Center/California Institute of Technology, funded by the National
Aeronautics and Space Administration and the National Science
Foundation.
AI, GT and ST acknowledge partial financial support
from the Agenzia Spaziale Italiana under contract ASI-INAF I/023/05/0.
\end{acknowledgements}

\bibliographystyle{}

\begin{thebibliography}{}

\bibitem[Belsole et al.(2003)]{Bel03} Belsole, E., Sauvageot, 
J.-L., Ponman, T.~J., \& Bourdin, H.\ 2003, \aap, 398, 1

\bibitem[Barnes(1985)]{Barnes85} Barnes, J.\ 1985, \mnras, 215, 
517

\bibitem[Barnes(1989)]{Barnes89} Barnes, J.~E.\ 1989, \nat, 338, 
123

\bibitem[von Benda-Beckmann et al.(2007)]{Benda}
von Benda-Beckmann, A.M. et al. 2007, astroph-0710.1297

\bibitem[Bode et al.(1993)]{Bodeetal} Bode, P.~W., Cohn, H.~N., 
\& Lugger, P.~M.\ 1993, \apj, 416, 17

\bibitem[Carter \& Read(2007)]{carterread} Carter, J.~A., \& Read, 
A.~M.\ 2007, \aap, 464, 1155

\bibitem[Cash(1979)]{cash} Cash, W., 1979, ApJ, 228, 939.

\bibitem[Colbert et al.(2002)]{colbert} Colbert, E.~J.~M., 
Weaver, K.~A., Krolik, J.~H., Mulchaey, J.~S., \& Mushotzky, R.~F.\ 2002, 
\apj, 581, 182

\bibitem[Coziol et al.(2000)]{coziol} Coziol, R., Iovino, A., 
\& de Carvalho, R.~R.\ 2000, \aj, 120, 47 

\bibitem[Coziol et al.(2004)]{coziol04} Coziol, R., Brinks, E., 
\& Bravo-Alfaro, H.\ 2004, \aj, 128, 68 

\bibitem[Coziol \& Plauchu-Frayn(2007)]{CoziolPlauchu-Frayn} Coziol, R., \& 
Plauchu-Frayn, I.\ 2007, \aj, 133, 2630  


\bibitem[Da Rocha and Mendes de Oliveira(2005)]{DaRochaMendes05}
Da Rocha, C. \& Mendes de Oliveira, C. 2005, \mnras, 364, 1069

\bibitem[Dickey \& Lockman(1990)]{dickey} Dickey, John M. \& Lockman, Felix
J., 1990, ARA\&A  28, 215.

\bibitem[Dos Santos \& Mamon(1999)]{dossantos} Dos Santos, S., \& 
Mamon, G.~A.\ 1999, \aap, 352, 1

\bibitem[Fukazawa et al.(2002)]{Fukazawa} Fukazawa, Y., Kawano, 
N., Ohta, A., \& Mizusawa, H.\ 2002, \pasj, 54, 527

\bibitem[G\'omez-Flechoso \& Domínguez-Tenreiro(2001)]{gomez} 
G\'omez-Flechoso, M. A., \& Domínguez-Tenreiro, R. 2001, ApJ, 549, L187

\bibitem[Heckman et al.(2005)]{heckman} Heckman, T.~M., Ptak, 
A., Hornschemeier, A., \& Kauffmann, G.\ 2005, \apj, 634, 161

\bibitem[Helsdon \& Ponman(2000)]{HelsdonPonman} 

\bibitem[Helsdon et al.(2001)]{Helsdon01} Helsdon S.F., Ponman T.J.,
O'Sullivan, E., Forbes, D.A, 2001, \mnras, 325, 693


\bibitem[Helsdon et al.(2005)]{Helsdon05} Helsdon S.F., Ponman T.J.,
Mulchaey, J.S., 2005 

\bibitem[Hickson(1982)]{Hickson} Hickson, P.\ 1982, \apj, 255, 
382 

\bibitem[Hickson et al.(1992)]{Hickson92} Hickson, P., Mendes de 
Oliveira, C., Huchra, J.~P., \& Palumbo, G.~G.\ 1992, \apj, 399, 353

\bibitem[Iovino(2000)]{Iovino2000} Iovino, A.\ 2000, IAU 
Colloq.~174: Small Galaxy Groups, 209, 25

\bibitem[Iovino et al.(2003)]{Iovino03} Iovino, A., de Carvalho, 
R.~R., Gal, R.~R., Odewahn, S.~C., Lopes, P.~A.~A., Mahabal, A., \& 
Djorgovski, S.~G.\ 2003, \aj, 125, 1660 

\bibitem[Jansen et al.(2001)]{refxmm} Jansen, F., et al.\
2001, \aap, 365, L1

\bibitem[Jeltema et al.(2006)]{Jeltema} Jeltema, T.~E., Mulchaey, J.~S.,
Lubin, L.~M., Rosati, P., B\"ohringer, H.\ 2006, \apj, 649, 649

\bibitem[Jones et al.(2003)]{jones03} Jones, L.~R., Ponman, 
T.~J., Horton, A., Babul, A., Ebeling, H., \& Burke, D.~J.\ 2003, \mnras, 
343, 627 

\bibitem[Mamon(1986)]{Mamon} Mamon, G.~A.\ 1986, \apj, 307, 
426

\bibitem[Mendes de Oliveira and Hickson(1994)]{MendesHickson94}
Mendes de Oliveira, C.     \& Hickson, P. 1994, \apj, 427, 684

\bibitem[Mulchaey(2000)]{M00} Mulchaey J.S. 2000, ARAA, 38, 289 

\bibitem[Mulchaey et al.(2003)]{Mul03} Mulchaey, J. S., Davis, D. S., 
Mushotzky, R. F., Burstein, D., 2003, ApJS, 145, 39

\bibitem[Mulchaey et al.(2006)]{Mulchaey06} Mulchaey, J. S., Lubin, L. M.,
Fassnacht, C., Rosati, P., \& Jeltema, T. E. 2006, \apj, 646, 133

\bibitem[Nandra et al.(2007)]{Nandra} Nandra, K., O'Neill, 
P.~M., George, I.~M., \& Reeves, J.~N.\ 2007, \mnras, 382, 194 

\bibitem[Osmond \& Ponman(2004)]{OsmondPonman2004} Osmond, J.P.F. 
\& Ponman, T.J. 2004
 2004 MNRAS, 350, 1511

\bibitem[Ota et al.(2004)]{Ota04} Ota, N., Morita, U., Kitayama, T.,
Ohashi, T. 2004, PASJ 56, 753


\bibitem[Panessa et al.(2006)]{panessa} Panessa, F., Bassani, 
L., Cappi, M., Dadina, M., Barcons, X., Carrera, F.~J., Ho, L.~C., \& 
Iwasawa, K.\ 2006, \aap, 455, 173

\bibitem[Pompei et al.(2007)]{Pompei07} Pompei, E., Dahlem, M., 
\& Iovino, A.\ 2007, \aap, 473, 399

\bibitem[Ponman et al.(1994)]{Ponman94} Ponman, T. J., Allen, D. J.,
Jones, L. R., Merrifield, M., McHardy, I. M., Lehto,
H. J., \& Luppino, G. A. 1994, Nature, 369, 462

 \bibitem[Ponman et al.(1996)]{Ponman} Ponman, T.~J., Bourner, 
P.~D.~J., Ebeling, H., \& Bohringer, H.\ 1996, \mnras, 283, 690

\bibitem[Rasmussen et al.(2003)]{rasmussen} Rasmussen, A., Kahn, S. M.,
\& Paerels, F. 2003, in The IGM/Galaxy Connection, ed. J. L. Rosenberg \&
M. E. Putman (Dordrecht: Kluwer), 109

\bibitem[Revnivtsev et al.(2007)]{revnivtsev} Revnivtsev, M.,  Churazov, E., Sazonov, S., Forman, W.,
Jones, C., 2007, astrp-ph 0702578

\bibitem[Ribeiro et al.(1996)]{ribeiro} Ribeiro, Andre L. B., de Carvalho, Reinaldo R., Coziol, Roger;
Capelato, Hugo V., Zepf, Stephen E. 1996, ApJ, 463, L5

\bibitem[Rose(1977)]{Rose77} Rose, J.~A.\ 1977, \apj, 211, 311 

\bibitem[Saracco \& Ciliegi(1995)]{SaraccoCiliegi} Saracco, P., \& 
Ciliegi, P.\ 1995, \aap, 301, 348 

\bibitem[Skrutskie et al.(2006)]{Skrutskie2006} Skrutskie,
M.~F., et al.\ 2006, \aj, 131, 1163

\bibitem[Strickland et al.(2004)]{Strickland} Strickland, D.~K., 
Heckman, T.~M., Colbert, E.~J.~M., Hoopes, C.~G., \& Weaver, K.~A.\ 2004, 
\apjs, 151, 193

\bibitem[Temporin et al.(2005)]{Temporin} Temporin, S., Ciroi, 
S., Iovino, A., Pompei, E., Radovich, M., \& Rafanelli, P.\ 2005, 
Starbursts: From 30 Doradus to Lyman Break Galaxies, Ap\&SS Lib. 329, 78

\bibitem[Trinchieri et al.(2003)]{Trinchieri2003} Trinchieri, G., 
Sulentic, J., Breitschwerdt, D., \& Pietsch, W.\ 2003, \aap, 401,

\bibitem[Trinchieri et al.(2005)]{Trinchieri2005} Trinchieri, G., 
Sulentic, J., Pietsch, W., \& Breitschwerdt, D.\ 2005, \aap, 444, 697 

\bibitem[Trinchieri et al.(2007)]{Trinchieri2007} Trinchieri, G., 
Breitschwerdt, D., Pietsch, W., Sulentic, J., \& Wolter, A.\ 2007, \aap, 
463, 153

\bibitem[T{\"u}llmann et al.(2006)]{Tuellmann} T{\"u}llmann, R., 
Pietsch, W., Rossa, J., Breitschwerdt, D., \& Dettmar, R.-J.\ 2006, \aap, 
448, 43 

\bibitem[Turner et al.(2001)]{refmos} Turner, M.~J.~L., et
al.\ 2001, \aap, 365, L27

\bibitem[Verdes-Montenegro et al.(1998)]{Verdesetal98}Verdes-Montenegro, 
L.,
Yun, M. S., Perea, J., Del Olmo, A., Ho, P. T. P. 1998, \apj, 497, 89

\bibitem[Verdes-Montenegro et al.(2001)]{Verdes01}Verdes-Montenegro, L.,
Yun, M. S., Williams, B. A., Huchtmeier, W. K.,
Del Olmo, A., Perea, J. 2001, \aap, 377, 812

\bibitem[Vikhlinin et al.(1999)]{Vikhlinin99} Vikhlinin, A., McNamara, 
B. R.,
Hornstrup, A., Quintana, H., Forman, W.,
Jones, C., \& Way, M., 1999, \apjl, 520, L1

\bibitem[White et al.(2003)]{hcg90ref} White, P.~M., Bothun, G., 
Guerrero, M.~A., West, M.~J., \& Barkhouse, W.~A.\ 2003, \apj, 585, 739 

\bibitem[Wilms et al.(2000)]{Wilms} Wilms, J., Allen, A., \& 
McCray, R.\ 2000, \apj, 542, 914 


\end{thebibliography}

 \end{document}